\RequirePackage{tikz}
\documentclass[twocolumn,trackchanges]{aastex631} 

\usepackage{newtxtext,newtxmath}

\usepackage[T1]{fontenc}

\DeclareRobustCommand{\VAN}[3]{#2}
\let\VANthebibliography\thebibliography
\def\thebibliography{\DeclareRobustCommand{\VAN}[3]{##3}\VANthebibliography}

\usepackage{slashbox}
\usepackage{graphicx}	
\usepackage{amsmath}	
\usepackage{xcolor}
\usepackage{multirow}
\usepackage{array}
\usepackage{comment}

\usetikzlibrary{patterns,decorations.markings,backgrounds}
\usepackage{pgfplots}
    \usetikzlibrary{intersections}
    \pgfplotsset{compat=1.11}
\pgfplotsset{
    every axis plot/.append style = {font = \normalsize}
  }
  
\newcolumntype{x}[1]{wc{#1}}
\newcolumntype{M}[1]{>{\centering\arraybackslash}m{#1}}

\newcommand{\solarmass}{\,{\rm M}_\odot}
\newcommand{\mkred}{\color{red}}
\newcommand{\mkblue}{\color{blue}}

\definecolor{slateblue}{rgb}{0.42,0.35,0.8}
\definecolor{ao(english)}{rgb}{0.0, 0.5, 0.0}
\definecolor{cadmiumgreen}{rgb}{0.0, 0.42, 0.24}

\usepackage{mathtools,amssymb,lipsum}

\usepackage{cuted}
\shorttitle{Mass determination method using multi-messenger measurements}
\shortauthors{Li et al}

\begin{document}

\title{A multi-messenger mass determination method for {LISA} Neutron-Star–White-Dwarf Binaries}

\correspondingauthor{ Jane SiNan Long (JSL),
Kaye Jiale Li (KJL), Albert, K. H. Kong (AKHK)}
\email{sinan.long.23@ucl.ac.uk (JSL), 
  j-li.19@ucl.ac.uk (KJL),
  akong@gapp.nthu.edu.tw (AKHK)}

\author[0000-0002-1657-0265]{Kaye Jiale Li}
\affiliation{
Mullard Space Science Laboratory, University College London, 
  Holmbury St Mary, Surrey, RH5 6NT, United Kingdom} 
  
\author[0000-0002-2126-0050]{Jane SiNan Long} 
\thanks{Present Address: Mullard Space Science Laboratory, UCL}
\affiliation{
Institute of Astronomy, 
National Tsing Hua University, 
  Hsinchu 30013, 
  Taiwan (ROC)} 

\author[0000-0002-7568-8765]{Kinwah Wu}
\affiliation{
Mullard Space Science Laboratory, University College London, 
  Holmbury St Mary, Surrey, RH5 6NT, United Kingdom} 
\affiliation{
Institute of Astronomy, National Tsing Hua University, 
  Hsinchu 30013, Taiwan (ROC)}   
\author[0000-0002-5105-344X]{Albert K. H. Kong}
\affiliation{
Institute of Astronomy, National Tsing Hua University, 
  Hsinchu 30013, Taiwan (ROC)}




\begin{abstract}

Determining the mass of the neutron stars   (NSs) 
  accurately improves our understanding of the   NS 
  interior and complicated binary evolution. 
However, the masses of the systems are 
  degenerate with orbital inclination angle when using solely 
    gravitational waves (GWs)
  or electromagnetic measurements, especially for face-on binaries. 
Taking advantages of both   GWs and 
  optical observations for {LISA} neutron-star–white-dwarf 
  (NS-WD) binaries, 
  we propose a mass determination method 
  utilising multi-messenger observational information. 
By combining the binary mass function 
  obtained from optical observations 
  and a   GW mass function, that we introduce,  
  derived from   GW observations, 
  we demonstrate how we can set improved constraints
  on the   NS mass and
  break the degeneracy in the 
  mass and viewing inclination determination.
We further comment on the universal relation 
  of the error bar of   the GW mass function versus 
  GW signal-to-noise ratio (SNR),
  and propose a simple method for the estimate of capability of 
  GW observations on mass determination with {LISA}. 
We show that for ultra-compact   NS-WD binaries
  within our Galaxy, the mass of the   NS 
  can be constrained to within 
  an accuracy of $\pm 0.2 \solarmass$ with the proposed method.





\end{abstract}




\section{Introduction}















%


Neutron stars (NSs) 
  are composed of the densest matter known to date, 
  and their mass density  
  is inaccessible in laboratory experiments.  
The masses of NS are around $1-2 \solarmass$,    
  and their upper limit is set 
  by the ability of the forces of quantum nature 
  to counteract the NS's self-gravity.  
Observations have 
  revealed several massive  ($\gtrsim 2 \solarmass $) NSs in 
  the neutron-star-white-dwarf (NS-WD) binaries 
\citep{Demorest2010Natur,Antoniadis2013Sci,
Cromartie2020,Fonseca2021ApJ}. 
 These observations have challenged 
  the existence of exotic matter, 
  which typically softens the equation of state (EOS) 
  compared to pure nucleonic matter, 
  thereby lowering the maximum mass a NS can support \citep{Baldo2000,Vidana2011,Moshfegh2013}.
A large parameter space for such exotic matter
  has been ruled out or at least disfavored \citep[see e.g.][]{Demorest2010Natur,Pang2021}.
However, models incorporating a condensed quark core remain viable \citep{Tang2021,Li2021ApJ,Akmal1998PhRvC,Alford2013PhRvD},
  and factors such as many-body corrections to 
  baryon-baryon interactions \citep{Akmal1998,Zhou2004,Li2008,Mojarrad2016}, 
  magnetic fields \citep{Zuraiq2023}, differential rotation \citep{Espino2019},  and finite temperatures \citep{Lattimer1991NuPhA} could alter the upper mass limit.
Firmly establishing the mass values for the heaviest NSs
  will tighten the constraints 
  on the allowed EOS of dense matter 
  and hence the structural models and 
  the internal compositions of the NS family 
  \citep{Lattimer2004Sci, Lattimer2007PhR}.
It also gives us a means 
  to study the behaviours 
  of complex phases 
  of dense quantum matter 
  under extreme pressure and gravity 
  \citep[][]{Avancini2008PhRvC,Baym2018RPPh,Tan2022PhRvD}.


Compact NS-WD binaries in our Galaxy 
  are candidate sources 
  for the space-borne gravitational wave (GW) detector 
  {LISA}. 
They could be formed through several channels.  
For instance, some may descend  
  from low-mass/intermediate-mass X-ray binaries 
  (LMXBs/IMXBs), 
  consisting of a NS accreting matter from 
  a low- or intermediate-mass giant  
  \citep[see e.g.][]{Tauris1999} 
  or even a degenerate WD companion 
  \citep[see e.g.][]{Tutukov1993}. 
They may also be products  
  of dynamical few-body interactions 
  occurring in very dense stellar environments, 
  e.g. cores of dense stellar clusters.  
The Galactic population of NS-WD binaries 
  is highly uncertain.  
It is estimated to range 
  from a few to a few hundreds 
  within {LISA} detection sensitivity,
  based on population synthesis models 
  or X-ray luminosity function 
  \citep[][]{Nelemans2001,Benacquista2001,Cooray2004,Chen2020,Korol2023}.   
The NS-WD binaries, 
  be they in the detached or semi-detached configuration, 
  with orbital evolution driven by 
  the angular momentum loss through gravitational radiation 
  \citep[see][]{Paczynski1971ARA&A}, 
  can provide 
  clean constraints 
  to the masses of component stars 
  through multi-messenger observations. 
The compact nature of the WD companion 
  allows the NS-WD binaries 
  to remain detached,  
  even when their orbital period, $P_{\rm orb}$,   
  evolves to as short as 
  about $\lesssim 10~{\rm min}$  
  \citep[][]{Tauris2018, Yu2021MNRAS, Chen2021MNRAS, Chen2022ApJ}.  
These NS-WD binaries naturally emit GWs. 
The frequencies of their   GWs 
  are expected to be from 0.5~mHz to 3~mHz, 
  which fall within the detectable range 
  of the {LISA} 
  \citep[see][]{Amaro2017}.  
Multi-messenger observations of NS-WD binaries 
  in GW and electromagnetic (EM) waves 
  will give us the opportunities 
  to probe the physics 
  of high-density matter in strong gravity 
  in addition to enhancing 
   our understanding of the nature, 
   the dynamics and the origins of NS-WD binaries.



The focus of this study 
  is to derive tighter constraints
  on the masses of the NSs 
  in NS-WD binaries detectable by {LISA}.  
We specifically introduce 
  a   GW mass function, 
  which is constructed 
  using the information from GW observations. 
This is complementary to 
  the conventional binary mass function 
  which is derived  
  from optical photometry or spectroscopy 
  or other EM/photonic observations.  
We use the GW analysis pipeline 
  {\tt gbmcmc} \citep{ldasoft2020}
  to generate mock {LISA} data 
  for parameter estimations. 
Using the results 
  from the Markov Chain Monte Carlo (MCMC) sampling 
  and the two mass functions, 
  which are almost diagonal, 
  we can constrain the orbital inclinations   
  and hence establish the masses of the NSs. 
We organise the paper as follows. 
The methodology and 
  the procedures are presented in \S \ref{sec:method} 
  and the results in \S~\ref{sec:result}. 
We discuss our findings   
   in \S~\ref{sec:discussion}. 
We also comment on the astrophysics of the target binaries. 
A short summary is given in \S~\ref{sec:conclusion}.

\section{Method} %
\label{sec:method}


\subsection{SNR threshold for gravitational-wave sources}

Consider a binary system consisting 
  of a NS of $m_{\rm ns}$ and 
  a low-mass WD of $m_{\rm c}$ 
  orbiting around each other 
  with a period $P_{\rm orb}$.  
The orbital revolution of the binary 
  leads to the emission of GW, 
  and the GW is characterised by a frequency $f_{\rm GW} \equiv 2/ P_{\rm orb}$. 
The loss of orbital energy and angular momentum 
  caused by the GW emission 
  bring the stars in the binary closer, 
  which in turns shortens $P_{\rm orb}$, 
  and increases $f_{\rm GW}$.
The rate of change in $f_{\rm GW}$ is given by 
\begin{align}
\dot{f}_{\rm GW} & = \frac{96}{5} \frac{G^{5/3} }{c^5} \pi^{8/3} \mathcal{M}^{5/3} 
{f_{\rm GW}}^{11/3}   \nonumber \\
& \approx 1.8 \times 10^{-11} \bigg( \frac{\eta}{0.3} \bigg) \bigg( \frac{M}{2.6 \solarmass}\bigg)^{5/3}
\bigg( \frac{1\,{\rm hr}}{P_{\rm orb} } \bigg)^{11/3} {\rm Hz}\, {\rm yr}^{-1}  
\end{align}  
\citep[see e.g.][]{Maggiore2008}, 
    where $M = m_{\rm ns} + m_{\rm c}$ 
  is the total mass of the binary, 
  $\eta =  (m_{\rm c} m_{\rm ns})/(m_{\rm ns} + m_{\rm c})^2$ is the symmetric mass ratio,  and 
\begin{align} 
\label{eq:chirpmass}
\mathcal{M} 
\equiv 
\frac{(m_{\rm c} m_{\rm ns})^{3/5}}{(m_{\rm ns} 
+ m_{\rm c})^{1/5}} \ ,
\end{align} 
is the chirp mass. 

For the parameters of interest in this study, 
  the drift in frequency is about 
  $\Delta f_{\rm GW} \approx 1 \times 10^{-11} {\rm Hz}$ 
  over a period of $4\,{\rm yr}$ (expected {LISA} mission lifetime). 
This value is much smaller than $f_{\rm GW}$.  
The GW from the system 
  is therefore practically quasi-monochromatic. 
Following \citet{Finn2000}, we define 
  the characteristic amplitude of the GW as  
  $h_{\rm c} \equiv h_{\rm o} \sqrt{2 f^2 /\dot{f}}$, 
  where $h_{\rm o} \equiv  \sqrt{ \left< {h_{+}}^2 + {h_{\times}}^2 \right> } = I(\iota) \sqrt{G \dot{E}/c^3}/(\pi d f)$ 
  is the rms amplitude
  of the two polarisations averaged 
  over one GW period.
The prefactor $I(\iota)$
  accounting for the effect of the inclination angle,  
  which is the rms of the angular part of $h_{+}$ and $h_{\times}$, 
  is given by  
\begin{align} 
\label{eq:IiotaDefinition}
I(\iota) \equiv  \sqrt{\frac{5}{4}} 
  \sqrt{\cos^2 \iota + \left( \frac{1+ \cos^2 \iota}{2}\right)^2 } \ ,   
\end{align} 
which is normalised such that $\int ({{\rm d} \Omega}/{4\pi}) I^2(\iota) = 1$. 
The signal-to-noise ratio (SNR) of this quasi-monochromatic GW is  
\begin{align} 
\label{eq:SNRdef} 
{\rm SNR}  & =  \sqrt{ \int_{-\infty}^{\infty} {\rm d}(\ln f) 
 \left(  \frac{h_{\rm c}(f) }{h_{\rm n}(f)} \right)^2}
\approx  \frac{\sqrt{2 f T } \, h_{\rm o} }{h_{\rm n}(f) }   
\end{align}  
 \citep{Flanagan1998,Finn2000,Moore2015,Robson2019}, 
where $h_{\rm n}(f) \equiv \sqrt{f S_{\rm n}(f)} $ is the sky-averaged rms noise 
  at frequency $f$, and $T$ is the observational time. 
$S_{\rm n}(f)$ is the one-sided, sky-averaged power spectral density (PSD) of the detector at frequency $f$.

We may take the sensitivity curve as  
\begin{align} 
\label{eq:Snf}
S_{\rm n}(f) & =  \frac{10}{3 L^2}\left\{ P_{\mathrm{OMS}}+2\left[1+\cos ^2\left(\frac{f}{f_*}\right)\right] \frac{P_{\mathrm{acc}}}{(2 \pi f)^4}\right\}  
 \nonumber \\ 
 & \hspace*{1.5cm}
 \times 
 \left[1+\frac{6}{10}\left(\frac{f}{f_*}\right)^2\right]  
\end{align}  
   \citep{Robson2019}.  
Here, $L= 2.5 {\rm ~Gm}$ 
  and $f_* = c/(2 \pi L) = 19.09{\rm ~mHz}$.
The single test mass acceleration noise single-link optical metrology noise $P_{\mathrm{acc}}$ and $P_{\mathrm{OMS}}$ are then 
\begin{equation} 
\left\{ 
\begin{aligned} 
&
P_{\mathrm{acc}}  = 9 \times 10^{-30} 
\left[ 1+\left(\frac{0.4 \mathrm{~mHz}}{f}\right)^2\right]
\left[ 1+\left(\frac{f}{8 \mathrm{~mHz}} \right)^4 \right]
~\frac{\rm m^2}{{\rm Hz}~{\rm s}^4} \ ;  \\
& P_{\mathrm{OMS}}  = 2.25 \times 10^{-22} 
\left[ 1+\left(\frac{2\mathrm{~mHz}}{f}\right)^4\right]
~\frac{\rm m^2}{{\rm Hz}} \ , 
\end{aligned} 
\right.
\end{equation}  
  so to be consistent 
  with the {\tt gbmcmc} package 
  used in this study\footnote{The 
  noise parameters are adapted from LISA LDC git:\url{https://gitlab.in2p3.fr/LISA/LDC/-/blob/master/ldc/lisa/noise/noise.py}}. 


Based on the formula of SNR in Eq.~\ref{eq:SNRdef},
  we define an effective GW strain
  $h_{\rm e} = \sqrt{2 f T } \, h_{\rm o} =  \sqrt{2 \mathcal{N}_{\rm cyc}} \, h_{\rm o}$,
  where $\mathcal{N}_{\rm cyc}$ is
  the number of wave cycles detected 
  within observational time $T$, 
  which is the same as that adopted in \cite{Chen2020}. 
The effective GW strain is then  
\begin{align} 
\label{eq:heffective}
h_{\rm e} & = 
  I(\iota) \sqrt{2 \mathcal{N}_{\rm cyc}} \frac{1}{\pi d f } \sqrt{\frac{G \dot{E}}{c^3}} \nonumber  \\ 
  & = 
 I(\iota) \frac{8}{\sqrt{5}} \frac{G^{5/3}}{d c^4} \pi^{2/3}
\mathcal{M}^{5/3} f^{2/3} \sqrt{ \mathcal{N}_{\rm cyc}} \nonumber   \\
& \approx 3 \times 10^{-20} \  I(\iota) \   
\bigg( \frac{30\,{\rm min}}{P_{\rm orb} } \bigg)^{7/6}    
 \bigg( \frac{\mathcal{M}}{ (2/5)^{1/5} \solarmass} \bigg)^{5/3} 
 \nonumber 
\\
& \hspace*{6em} \times 
\bigg( \frac{T}{4\,{\rm yr}} \bigg)^{1/2}
\bigg(\frac{10\,{\rm kpc}}{d}  \bigg) \ , 
\end{align} 
  and the SNR can be visually determined 
  by how far this effective strain is above the noise amplitude $h_{\rm n}(f) $. 

While it is conventional to set ${\rm SNR}=5$ as the criterion of detection,
  this threshold is merely representative. 
Depending on the (time-delay interferometry) TDI channels considered,
  whether sky-averaged has been performed and how is it performed, 
  the amplitude of the Galactic confusion noise, 
  and whether the orbital orientation is averaged out,
  the SNR value of the exact same system could differ by a factor of a few.
For example, the package 
    {\tt gbmcmc} 
    \citep{ldasoft2020} 
   has a more detailed definition of SNR 
   \citep[see][]{Littenberg2019} 
using a combination of {\it A} and {\it E} channel
  with realistic responses to different sky locations, polarization, and
  inclination angles. 
For simplicity, we have considered two {\it X}-channels
  and averaged over the source's sky position and detector orientation angle.
For the system we considered in this work,
  our SNR definition in Eq.~\ref{eq:SNRdef} is consistent in number with 
  SNR(gbmcmc), with a difference of $\le 20\%$, mainly depending on the sky position. 
%
To account for this subtle difference in definition across literature,
  we relax this criterion to $\rm SNR=2$ and consider a wider range of systems, 
  including those with $2<{\rm SNR}<5$. 
Nevertheless, the SNR (or sometimes ${\rm SNR(gbmcmc)}$ 
  as calculated by the {\tt gbmcmc} package) 
  of each system will be listed to indicate the possibility of it being detected.

\begin{figure}
	\includegraphics[width=\columnwidth]{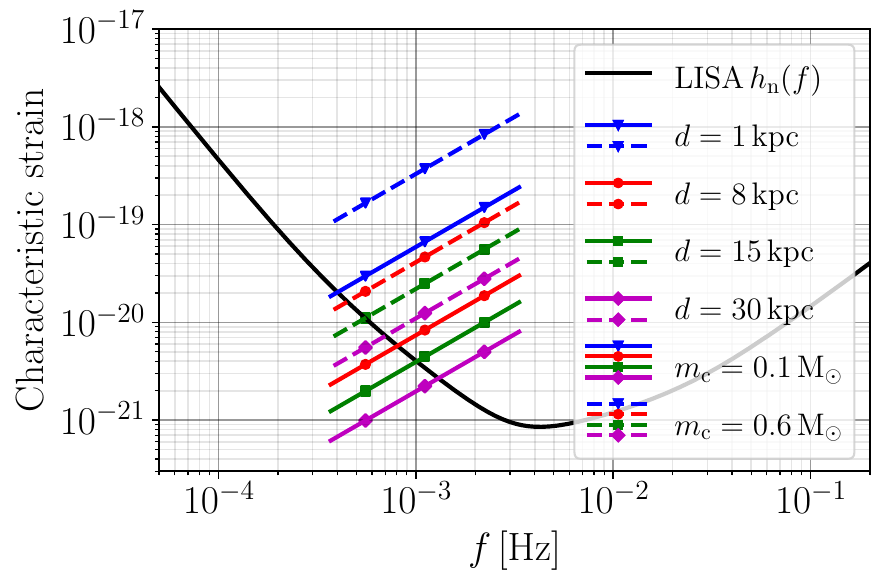} \vspace*{-0.5cm}
    \caption{The range of inclination angle-averaged
    effective GW strain of the binary system 
       with $m_{\rm ns}=2.0\;\! \solarmass$  
       and $m_{\rm c}=0.1,0.6 \;\! \solarmass$
       at $d=1$, $8$, $15$, and $30\;\!{\rm kpc}$.  
Each line represents 
  a specific type of binary system
  of the same $m_{\rm c}$ and $d$,
  with orbital period $P_{\rm orb}$ spanning from
  $10\,{\rm min}$ to $90\,{\rm min}$.
The three nodes are marked at 
  $P_{\rm orb}=60, \ 30, \ 15 \ {\rm min}$ from left to right respectively.
The integration time of the observations 
  is set to be 4 years. 
{LISA} sensitivity curve is defined in Eq.~\ref{eq:Snf}.}
    \label{fig:TidarrenRange}
\end{figure}

In this study, 
  we consider NS-WD binaries with orbital period 
  $P_{\rm orb} = 10-60\,{\rm min}$,
  which is within LISA's sensitivity range . 
The parameter choice extends toward lower frequencies,
  as systems with longer periods have longer lifetimes compared to more compact systems,
  thus dominating the population.
The distance range for these systems is 
  chosen to be $d=1-30\,{\rm kpc}$, 
  covering our entire   Galaxy. 
In particular, we   choose hypothetical systems 
  very close by to study parameter degeneracy and 
  demonstrate the universality of our findings for high SNR systems.
The existence of such   systems within our 
    Galaxy is discussed in Sec.~\ref{Existence}.
Without losing generality 
  we assume that the binary has a circular orbit.  
The mass of the NS is $2 \solarmass$\footnote{The existence of $2 \solarmass$ NSs is supported by 
  observations of NS-WD binaries \citep{Demorest2010Natur,Antoniadis2013Sci,Cromartie2020,Fonseca2021ApJ}
  which motivates our choice of this value.
In the literature, $2 \solarmass$ is often used as a benchmark to
  identify the existence and properties of exotic matter \citep{Lattimer2010,Akmal1998}.
However, exotic matter can still support NS exceeding $2 \solarmass$,
  with a critical threshold likely a few tenths of a solar mass higher
  -- though this remains highly model-dependent and complex \citep[see e.g.][]{Char2014,Zuraiq2023}. 
For simplicity, we adopt this convention while 
  noting that our main result is insensitive to the specific choice.
}
  and the mass of the companion star is 
  within the range $m_{\rm c}=0.1-0.6\solarmass$,
  although the majority of our results 
  can be extended to a wider parameter range. 
Fig.~\ref{fig:TidarrenRange} 
  illustrates the effective GW strains of the NS-WD binaries,  
  for a range of distances, orbital periods, and companion masses.  
The {LISA} sensitivity curve is included in the graph for comparison.


\subsection{Parameter estimation method}
\label{sec:parameterestimation}
NS-WD binaries have the potential to be identified through 
  EM observations before {LISA} is launched, 
  contributing to the family of
  verification binaries for {LISA} \citep{Stroeer2006,Korol2017,Kupfer2018,Burdge2019Natur,Burdge2019,Burdge2020ApJ,Kilic2021,Johnson2023MNRAS, Finch2023,Kupfer2023}.  
Alternatively, if the WD companion is too faint for 
  detection in optical surveys, GW observations could 
  offer localisation information for the binary, 
  enabling targeted follow-up searches. 
In this study, we assume that the binary has been 
  identified through optical observations, 
  either before or after GW identification.
  Through spectroscopic analysis, supplemented with photometric analysis, the orbital periods $P_{\rm orb}$ and phase resolved radial velocities $K$ of the WD are measured and are then used to construct the binary mass function,
\begin{align} 
\label{eq:massfunction}
f(m)  = \frac{P_{\rm orb} K^3}{2 \pi G}  
  = \frac{{m_{\rm ns}}^3 \sin^3\!\iota}{(m_{\rm ns} + m_{\rm c})^2}    \  . 
\end{align} 
  
Currently there has been no confirmed identifications 
  of our target compact (10$-$60 min) NS-WD binaries\footnote{The absolute brightness of white dwarfs 
    is $M_{V} \sim 11.5$ \citep[see e.g.][]{Mickaelian2022OAst}. 
   This corresponds to the visual magnitude $V \approx 26$ 
   at a distance of $8~{\rm kpc}$. 
 The 8-m class ESO Very Large Telescopes (VLT) 
   are capable of observing source of $V \approx 27$ 
   with sensible SNR at an integration time of 1 hr \citep[see][]{Reddy2019Icar}. 
 While observation for our target NS-WD binaries 
   at distance exceeding about $4~{\rm kpc}$ 
   with reliable photometric determination of periodic 
   variations is beyond 
   the capacity of current ground-based telescopes, 
   with the coming of the 30-m class telescopes, such as the Extremely Large Telescope 
     (ELT) and the Thirty Meter Telescope (TMT), 
     resolving 10~min photometric variations  
     at $V\approx 26$ would be achievable  
     for hours of integration time. 
The simple estimate above has not taken account 
  of important local effects, such as night-sky brightness 
  and atmospheric scintillations. 
All these would restrict 
  the size of observational volume 
  for the target binaries to reside. 
We may be able to circumvent these 
  in space-base observations, 
  where night-sky brightness, seeing, and other atmospheric effects  
  would not be present.}.
Estimating the uncertainty in determining 
  the binary mass function poses a non-trivial challenge. 
This uncertainty arises 
  from factors such as the data quality,   
  as well as
  the modeling the WD's brightness variations 
  due to tidal effects 
  and/or pulsar irradiation. 
Given these complexities, we find it impractical to 
  estimate the uncertainty directly 
  for an unspecified system 
  over a multi-dimension parameter space. 
Instead, focusing on GW observations, 
  we assign specific values to the binary mass function 
  uncertainty (i.e. $5-20\%$) to assess the sensitivity of mass estimates under these conditions.

\begin{figure*}
\vspace{-3.5em}
\centering
\includegraphics[width=1.8\columnwidth]{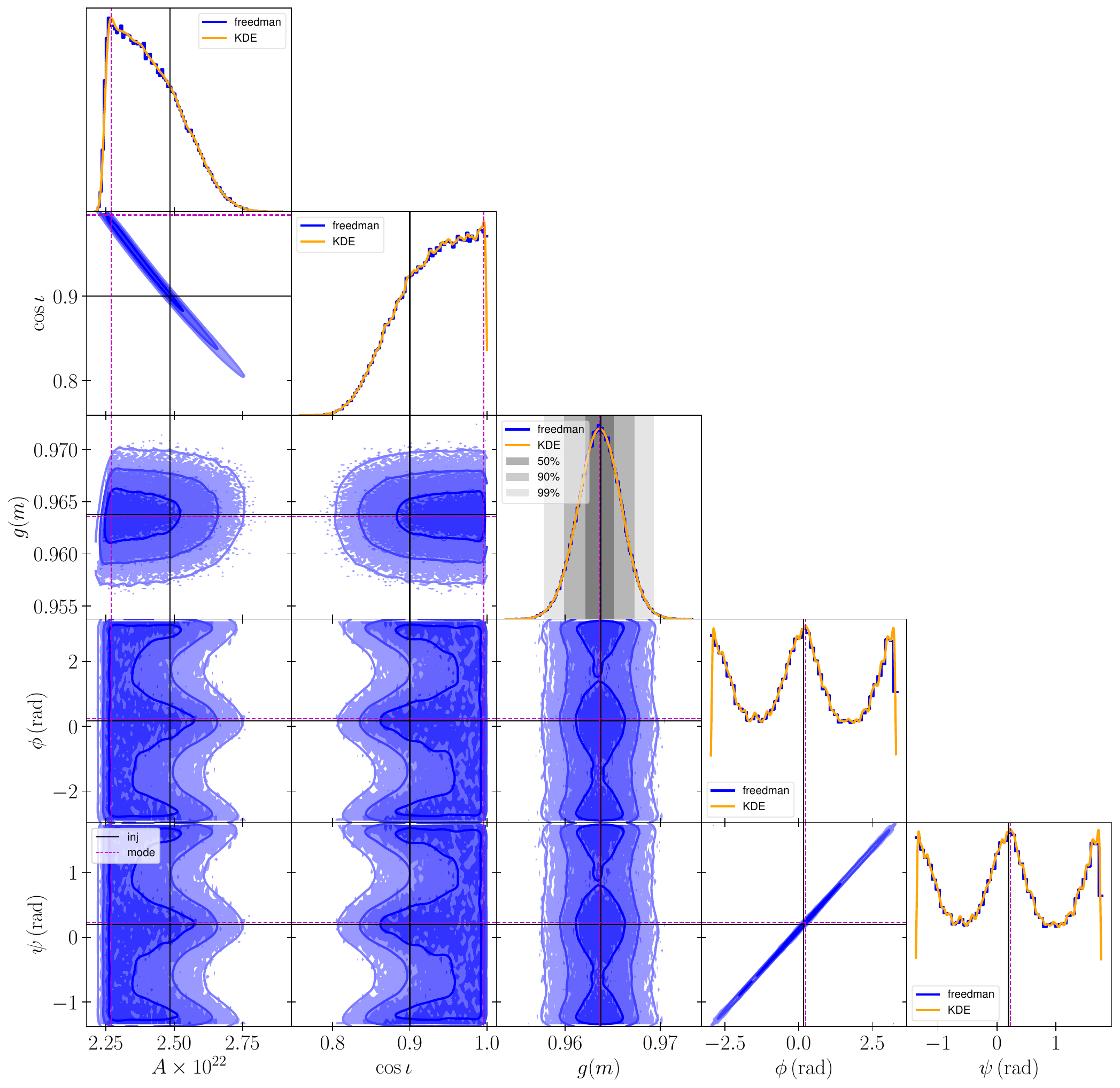}
\vspace{-1.0em}
\caption{The parameter estimation 
of the GW emitted by a binary system with 
 $m_{\rm c} = 0.5\solarmass$, $P_{\rm orb}=15\,{\rm min}$,
  and $\cos \iota = 0.9$   (nearly face-on) 
  located at $3\,{\rm kpc}$ away.
The detection time is $4$-year.
  The SNR of this system is $\approx 260$.
The contour lines represent the $50\%$, $90\%$, and $99\%$ confidence intervals. 
The histogram is calculated using the {\tt astropy} package \citep{Astropy2022ApJ}  and the kernel density estimation (KDE) is calculated by {\tt getdist} package \citep{Lewis2019arXiv}.}
\label{fig:Acosi-degeneracy}
\end{figure*}

For the GW part, we made use of the package {\tt gbmcmc} \citep{Littenberg2020} to perform mock parameter estimation of such binary systems with different inclination angles, orbital periods, and distances. 
Each binary is fitted with typically eight parameters including orbital frequency $f$, 
  frequency derivative $\dot{f}$,
  GW amplitude $A$, inclination angle $\iota$, 
  polarisation angle $\psi$, initial orbital phase $\phi$ 
  of GW and sky localization, where 
\begin{align} 
\label{eq:amplitude}
A & \equiv \frac{2}{d} \frac{G^{5/3}}{c^4}
\left(\pi f \right)^{2/3} \mathcal{M}^{5/3} \ . 
\end{align} 
 We have excluded noise in the mock data so as to 
  derive a clearer degeneracy between system parameters. 
In this analysis, 
  we have assumed such a system to be readily detected via EM observations 
  with known sky location and orbital frequency, and hence $f$.
The prior of polarisation angle and initial orbital 
  phase is uniform in $[0,\pi]$ and $[0,2\pi]$, respectively.
We restrict our study to prograde orbit
  and set the prior on the inclination angle such that $\cos \iota$ 
  is uniform in $[0,1]$. 
The prior amplitude $A$ is log-uniform between  
 $\ln A = -60$ and $\ln A = -45$.  

In the mock parameter estimation, the system is assumed to be located at 
  $\sin({\rm latitude})=0.6080$ and  ${\rm longitude}=2.9737$ 
  in solar system barycenter ecliptic coordinates 
  and it consists 
  of a NS with $2 \solarmass$
  and a companion with $m_{\rm c}=0.1-0.6\solarmass$,
  with an observational duration of the GW signal of $4\,{\rm yr}$. 
The system is assumed to have negligible accretion such that 
  $\dot{f} = \dot{f}_{\rm GW} \ll f/4\,{\rm yr}$
  and the source is essentially monochromatic.
In this study, we aim at demonstrating the capability 
  of GW observations, and hence have adopted a simplified model
  by assuming that the companion mass and binary distance 
  are known a priori from previous EM observations 
  (see Appendix~\ref{app:A} for a detailed discussion).
With the information of $m_{\rm c}$ and distance $d$ known,
  the mass of the NS can be exactly determined from a given $A$ by 
  combining Eq.~\ref{eq:chirpmass} and Eq.~\ref{eq:amplitude}. 
In practice, both the companion mass and the distance
  would be measured with an error bar. 
For such a monochromatic source,
  the distance is completely degenerate 
  with the chirp mass.
Therefore, the uncertainty of distance
  can be translated into the uncertainty of GW chirp mass
  via error propagation. 
  
The error propagation for 
  companion mass is more subtle as 
  it appears in both the binary mass function and GW amplitude. 
We will show how the error of companion mass
  affects the error of the NS's mass   in 
  appendix \ref{app:A}. 

  

While it is straightforward to calculate the system's chirp mass 
  from  GW amplitude $A$ using Eq.~\ref{eq:amplitude},
    $A$ is not necessarily 
  a good parameter due to its degeneracy with the inclination angle in GW detection. 
The two observed polarisation modes $h_{+} \propto A (1 + \cos^2 \iota)/2 $ 
  and $h_{\times} \propto A \cos \iota$,
  suggest that  a system with larger $A$ and larger inclination angles
 can be mistaken as a system with smaller $A$ and smaller inclination angles. 
Significant degeneracy appears even for signals with large SNR when the system is close to face-on.
For example, Fig.~\ref{fig:Acosi-degeneracy} shows 
  the parameter estimation 
  for an unrealistic injected signal emitted by a binary with 
  $m_{\rm c} = 0.5\solarmass$, 
  $P_{\rm orb}=15\,{\rm min}$,
  $\cos \iota = 0.9$ at a distance of $3\,{\rm kpc}$.
  This system has ${\rm SNR}\approx 260$
after a $4$-year integration time.
The parameter estimation for such 
  a luminous system is expected to be very precise,
  but as shown by the 1D marginalized 
  probability density function (PDF) of $A$ and $\cos \iota$,  
  both PDFs differ significantly from a bell-shaped distribution and 
  the locations of maximum posteriors mismatch the true values. 
This degeneracy motivates the definition of a new parameter that combines
  the information of both amplitude and inclination angle.

We define a   GW mass function from GW measurement,
  similar to the binary mass function from measurements of radial velocity variation (e.g. Eq.~\ref{eq:massfunction}), as 
\begin{align}
\label{eq:GWmassfunction}
 g(m) & =  \left[\frac{A d}{2(\pi f)^{2/3} }
  \frac{c^4}{G^{5/3}}
 \sqrt{\cos^2 \iota + \left( \frac{1+ \cos^2 \iota}{2}\right)^2}  \right]^{3/5} 
  \nonumber  \\
 & = \frac{(m_{\rm ns} m_{\rm c})^{3/5}}{(m_{\rm ns}+m_{\rm c})^{1/5}} \left[ \cos^2 \iota + \left( \frac{1 + \cos^2 \iota}{2} \right)^2 \right]^{3/10} \ .
\end{align}
Similar to that of the binary mass function, 
  the first equal sign shows how   the GW mass function
  is derived from direct observation parameters, while the second equal sign 
  indicates its relation with the intrinsic binary parameters.   
  Here, $A$ and $\cos \iota$ are obtained from 
  GW parameter estimation, while $f$ and $d$ 
  are known a priori from EM observations.
This   GW mass function is proportional to 
  $\left< {h_{+}}^2 + {h_{\times}}^2 \right>^{3/10}$,
  taking into account the fact that {LISA} 
  measures a combination of $h_{+}$ and $h_{\times}$ as it rotates 
    in space. 
As shown in the bottom-right panel of Fig.~\ref{fig:Acosi-degeneracy},
  the PDF of the   GW mass function has a nice symmetrical bell-shape,
    as opposed to the PDFs of $A$ and $\cos \iota$ which are not as meaningful when viewed individually. 

\begin{figure}
\centering
\includegraphics[width=1\columnwidth]{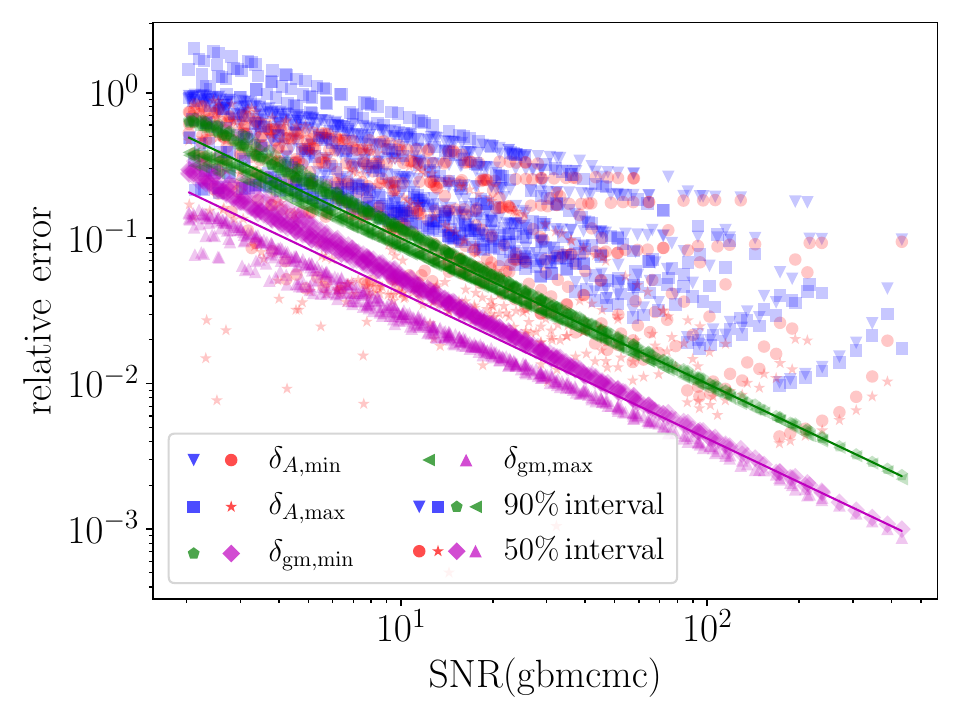}
\vspace*{-2.0em}
\caption{The relative error of the parameter 
  $A$ and $g(m)$ versus 
  the SNR of the system, at a confidence interval of $50 \%$ and $90 \%$ respectively. 
The upper and lower error bar of $g(m)$ is defined as 
  $\delta_{{\rm gm},\max} \equiv (\max g(m)-g(m)_0)/g(m)_0$,
  $\delta_{{\rm gm},\min} \equiv (g(m)_0-\min g(m))/g(m)_0$,
  and similarly for $A$. 
Some $\delta_{A,\max}$ values are missing as
  the strong degeneracy between $A$ and $\cos \iota$ has led to 
  $\max A < A_0$, as demonstrated in Fig.~\ref{fig:Acosi-degeneracy}. 
The relative errors of $g(m)$ roughly 
  follow power law relation with respect to SNR
  whereas the error bars of $A$ have large scatters. 
The solid lines represent the power-law relation reported in 
  Eq.~\ref{eq:gmErrorBar},
  which are derived by fitting the average of the maximum 
  and minimum relative errors. 
The data shown are calculated with $m_{\rm c} = 0.6\solarmass$
  and $P_{\rm orb} = 20-60\,{\rm min}$,
  although the results for other masses and orbital periods
  are very similar. 
}
\label{fig:gmAerrorbarBulk}
\end{figure}

The   GW mass function admits 
  advantages in the parameter estimation that
  involves multi-messenger observations regardless 
  of the orbital configuration of 
  the NS-WD binary. 
The accuracy of parameter estimation 
  of $g(m)$ follows a simple relation with the SNR of the binary system,
  particularly at high SNRs. 
Fig.~\ref{fig:gmAerrorbarBulk} compares 
  the accuracy on the parameter estimation 
  for $A$ and $g(m)$
  achievable for the NS-WD binaries. 
Compared with that of $A$, the relative size of 
  the confidence interval
  (i.e. the error bar) of $g(m)$
  roughly follows a power law relation with the SNRs of the system.
We define $g(m)_{\max,\min}= g(m)_0 (1 \pm \delta_{\rm gm})$,
  such that $g(m)_0$ represents the true value of the   GW mass function.
The averaged relative error of $\delta_{\rm gm}$
  is reasonably well fitted by the following relations: 
\begin{equation} 
\left\{  
\hspace*{0.2cm}
\begin{aligned} \label{eq:gmErrorBar}
\delta_{{\rm gm},50\%}  & \approx 0.042 \left( \frac{10}{\rm SNR} \right) \ ; \\
\delta_{{\rm gm},90\%}  & \approx 0.10 \left( \frac{10}{\rm SNR} \right) \ ,
\end{aligned} 
\right.
\end{equation}
  which are appropriate for NS-WD binaries with ${\rm SNR} \in [3,\ 400]$.
This relation has been validated across a range of companion masses,
  distances, inclination angles ($\cos \iota = 0.1-0.9$),
  orbital periods ($P_{\rm orb} = 10-60\,{\rm min}$) 
  and amplitudes ($A \in [3 \times 10^{-23} , \ 7 \times 10^{-22}]$),
  provided the ${\rm SNR}$ falls within this range. 
However, caution is advised when applying this relation,
  as elaborated at the end of this section.
As shown in Fig.~\ref{fig:gmAerrorbarBulk}, 
  the relative error of $g(m)$ is almost always smaller than
  that of $A$ at the same confidence interval.
For a nearly edge-on system with $\cos \iota \sim 0$, 
  the relative error bar of $g(m)$ is about $3/5$ of that of $A$,
  as a consequence of $g(m) \propto A^{3/5}$,
  except for when ${\rm SNR(gbmcmc)} \le 3$.
For a nearly face-on system, the relative error bar of $A$
  is much larger than that of $g(m)$, as a consequence of 
  the strong degeneracy between $A$ and $\cos \iota$ (see   Fig.~\ref{fig:Acosi-degeneracy}).

  Deviation from the simple power-law relation
  is anticipated at low SNR, 
  leading to a more stringent upper bound 
  $\delta_{{\rm gm},\max}$ and 
  a less restrictive lower bound $\delta_{{\rm gm},\min}$. 
This outcome is a consequence of the 
  log-uniform distribution of the prior for parameter $A$. 
  
For such low ${\rm SNR}$ systems, noise can significantly 
  distort or completely obscure 
  the power-law relation, highlighting a limitation 
  that requires further investigation.
Additionally, this relation assumes prior knowledge of 
  the system's orbital frequency, distance, 
  and sky location;
when these parameters are subject to measurement uncertainty, 
  non-trivial modifications to the relation Eq.~\ref{eq:gmErrorBar}
  are expected. 
Given the faintness of the WD, distance is likely the limiting factor. 
Fortunately, its effect on the   GW mass function error
  takes a simple form, as discussed in Appendix~\ref{app:A}.
The impact of uncertainty in 
  sky location and orbital frequency could be much
  more complicated, and we look forward to future study
  addressing this issue.   
Overall, the empirical relation in Eq.~\ref{eq:gmErrorBar} 
  provides a straightforward way to estimate the accuracy of 
  the   GW mass function achievable with LISA observations, 
  bypassing the need for mock parameter estimation.


\section{Results}  
\label{sec:result}
 
 \subsection{Analytical estimates of NS's mass by two mass functions}

As both optical observations and analysis of GW detection indicated, 
  the main issue for binary parameter estimation is the undetermined mass-inclination degeneracy.
For a binary system with both optical observations and GW detection,
  both mass functions can be measured, each subject to an error bar. 
As we have assumed that the companion mass $m_{\rm c}$ is known, 
  the maximum mass can be calculated by solving  
\begin{equation} 
\left\{ 
\hspace*{0.2cm}
\begin{aligned} \label{eq:mNSMax_fmgm}
f(m)_{\max} &  = \frac{({m_{{\rm ns},\max}})^3 \sin^3 \iota}{(m_{{\rm ns},\max} + m_{\rm c})^2} \ ; \\
g(m)_{\max} &  = \frac{(m_{{\rm ns},\max} m_{\rm c})^{3/5}}{(m_{{\rm ns},\max}+m_{\rm c})^{1/5}}
 \left[ \cos^2 \iota + \left( \frac{1 + \cos^2 \iota}{2} \right)^2 \right]^{3/10}  
\end{aligned} 
\right. 
\end{equation} 
simultaneously, and similarly for the minimum mass.
In addition to the joint constraint,
  the binary mass function places 
  a lower bound of $m_{\rm ns}$ via 
\begin{align} 
\label{eq:mNSMin_fm}
\frac{({m_{{\rm ns},\min}})^3}{(m_{{\rm ns},\min}+m_{\rm c})^2} \ge f(m)_{\min} \ .
\end{align}   
For an edge-on system, the variation of radial velocity is the maximum,
  leading to a stringent lower bound of the mass.
  The   GW mass function alone also provides an upper bound
\begin{align} 
\label{eq:mNSMax_gm}
\frac{(m_{{\rm ns},\max} m_{\rm c})^{3/5}}{(m_{{\rm ns},\max}+m_{\rm c})^{1/5}} \le
2^{3/5} g(m)_{\max} \ ,
\end{align} 
that is only relevant for high SNR systems 
  that are close to an edge-on configuration.
For a face-on system, we have $f(m)_{\min} \approx 0$,
  and the lower bound is trivial.

\begin{figure}
\centering
\includegraphics[width=1\columnwidth]{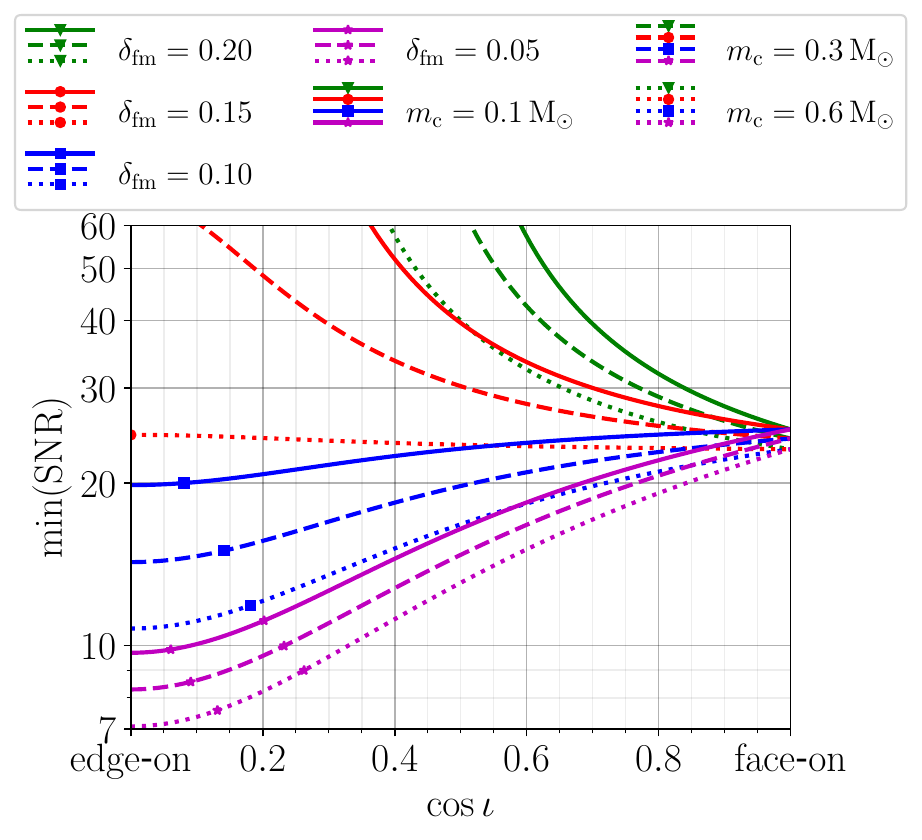}
\vspace*{-2.0em}
\caption{The minimum SNR of the NS-WD binaries required 
to achieve an measurement of $m_{\rm ns} = (2 \pm 0.2) \solarmass$
for different companion masses and different measurement accuracy 
  of $\delta_{\rm fm}$.
The marker(s) on each line represent the value of $\cos \iota$,
  below which the lower bound of $m_{\rm ns}$ placed by $f(m)_{\min}$ alone
  is equivalent or better than $1.8\;\! \solarmass$. 
  From right to left, the lower bound due to $f(m)_{\min}$ alone is
  $1.8$ and $1.9\;\! \solarmass$.
For some lines, some markers on the left are missing 
  when their best constraints
  can not reach $1.9\;\! \solarmass$ level.}
\label{fig:dm01}
\end{figure}

Using the empirical relation Eq.~\ref{eq:gmErrorBar}
  the minimum GW SNR required to achieve a measurement 
  of (at least) $m_{\rm ns} = (2 \pm 0.2) \solarmass$ at $90\%$ confidence level
  is shown in Fig.~\ref{fig:dm01}.
  
When $f(m)$ is measured accurately with $\delta_{\rm fm} \lesssim 10\%$,
  the minimum SNR required to probe $m_{\rm NS}$ with the same precision
  is smaller for nearly edge-on systems. 
If the binary mass function 
  is measured with $\pm 5\%$ accuracy or smaller,
  even systems with SNR $\lesssim 10$
  can constrain NS mass within
  $\pm 0.2\solarmass$ for optimal system configuration. 
The lower bound, 
  which is particularly important for 
  constraining NS internal structure, is usually 
  better than $0.2 \solarmass$ for nearly edge-on systems,
  primarily due to the constraint imposed by $f(m)_{\min}$ alone.
For nearly face-on systems, 
  the lower bound of the binary mass function $f(m)$ vanishes,
  and the joint constraint depends 
  mainly on the GW observations. 
  With such precise binary mass function measurements, 
  NS-WD binaries with SNR $\ge 25$ can determine 
  the mass to within $\pm 0.2\solarmass$, regardless of other orbital parameters.
When the binary mass function is measured less accurately 
  with $\delta_{\rm fm} \ge 10\%$,
  the preference for inclination angle is reversed. 
For nearly edge-on configuration, the constraint is primarily determined by $f(m)$. 
Therefore, when $f(m)$ is measured with large uncertainty, 
  even a perfect measurement of $g(m)$ 
  is insufficient to constrain $m_{\rm NS}$ to within $\pm 0.2\solarmass$.
However, it should be noted that if such a system has a sufficiently 
  high SNR, the $A-\cos\iota$ degeneracy disappears 
  (see e.g., Fig.~\ref{fig:fmgm}), 
  and the constraint from GW observations alone could be better than $\pm 0.2\solarmass$.
This is not captured by Fig.~\ref{fig:fmgm}, which is calculated using
  the   GW mass function via Eq.~\ref{eq:gmErrorBar}.

\begin{figure}
\centering
\includegraphics[width=1\columnwidth]{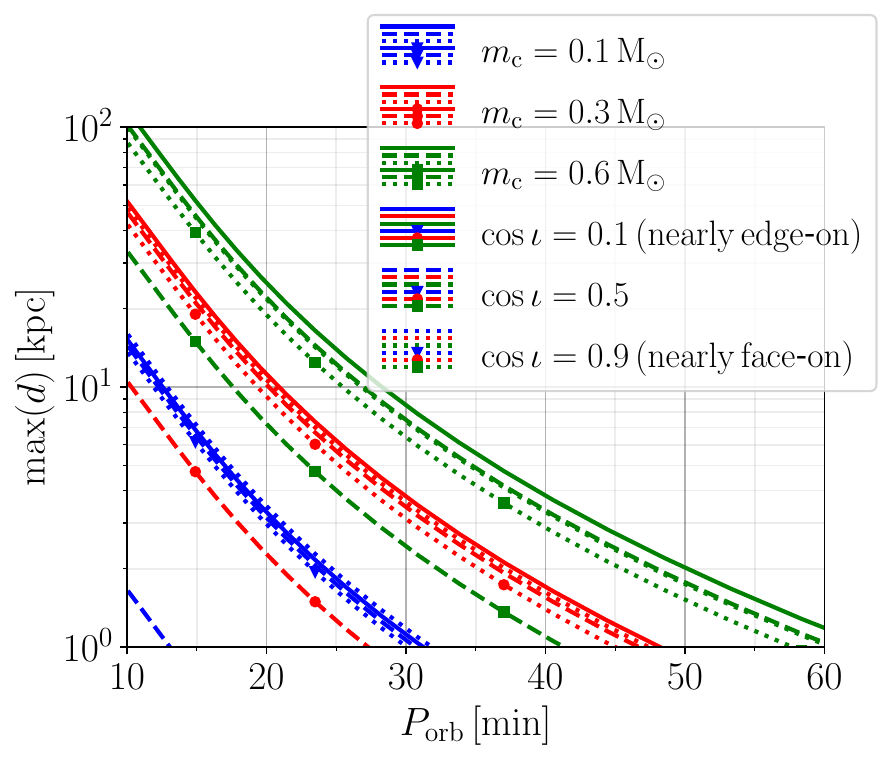}
\vspace*{-2.0em}
\caption{The maximum distance of the NS-WD binary
  for probing a measurement
  of $m_{\rm NS} = 2 \pm 0.2 \solarmass$. 
  The unmarked lines represent cases with $\delta_{\rm fm} = 5\%$
  and the marked lines represent cases with $\delta_{\rm fm} = 20\%$.
  It is important to highlight that, 
  for cases with $\cos \iota=0.1$   and $\delta_{\rm fm} = 5\%$, 
  the lower bound exceeds $1.8\solarmass$.}
\label{fig:dm02}
\end{figure}

Interestingly, the minimum SNR
  for binaries with a heavier companion
  is smaller than that for a lighter companion. 
This dependence becomes prominent when 
  $f(m)$ is measured with a $10\%$ error,
    in which case the minimum SNR for a heavy companion is 
  approximately half that of a lighter companion.
As NS-WD binaries with heavier WD has larger GW amplitude,
  this suggests that accurate NS mass measurement 
  prospects lie in such heavy NS-WD binaries. 
Using Eq.~\ref{eq:SNRdef},
  the minimum GW SNR can be translated into
  a relation between maximum binary distance 
  and orbital period. 
The result is shown in Fig.~\ref{fig:dm02},
  in which we adopt   $\delta_{\rm fm} = 5\%,\,20\%$
  and $\delta_{{\rm gm},90\%}$.
Due to the dependency of SNR and minimum SNR 
  on $m_{\rm c}$, 
  the probing distance for NS-WD binaries with a heavier 
  companion of 
  $m_{\rm c} = 0.6 \solarmass$ is   about ten times
  larger than that for a
  lighter companion with $m_{\rm c} = 0.1\solarmass$. 
Assuming a median distance of $10\,{\rm kpc}$,
  NS-WD binaries with 
  $P_{\rm orb} \lesssim 30\,{\rm min}$
  and a heavy companion 
  (i.e. $m_{\rm c} \gtrsim 0.6 \solarmass$)
  are the more promising candidates for probing NS mass.
  
The generic results to Eq.~\ref{eq:mNSMax_fmgm}
  is complicated in form,
  and for the ease of computation, we present
  an approximate formula in Appendix~\ref{app:A}.
  The formula is applicable when
  the relative errors 
  of all quantities are much smaller than unity.
Supplemented by the empirical relation outlined in Eq.~\ref{eq:gmErrorBar}, 
  the approximate formula
  can be employed to estimate the
  joint constraint on the NS mass for 
  NS-WD binaries with multimessenger method. 

\subsection{The joint constraints on the NS's mass}

\begin{figure*}
\centering
\includegraphics[width=2\columnwidth]{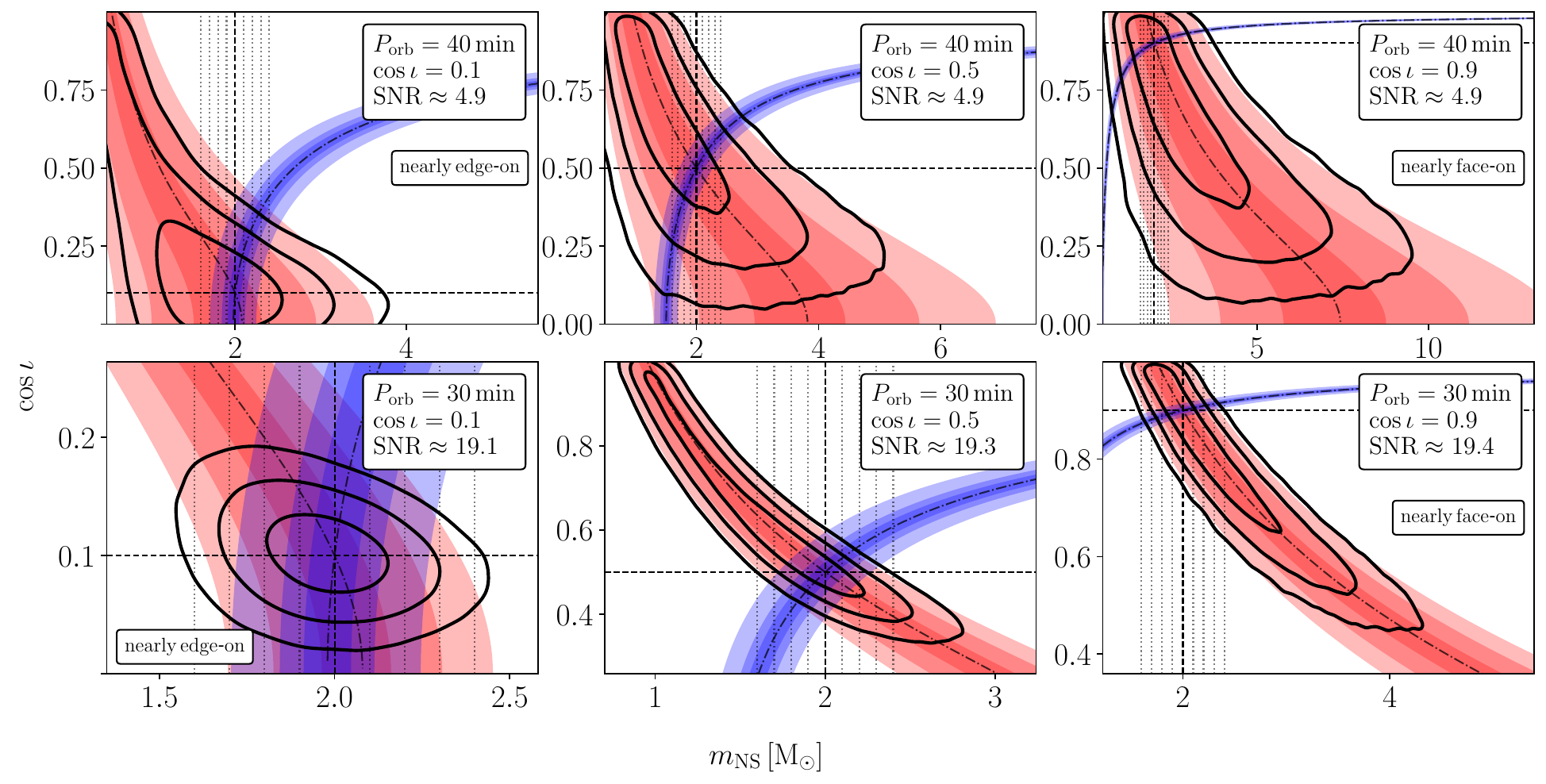}
\caption{
The constraints on the binary parameters for 6 different binary systems
  with $m_{\rm ns}=2\solarmass$ and $m_{\rm c}=0.6\solarmass$
  located at (top left to right) $d=5.22,8.11,12.94\,{\rm kpc}$
  and (bottom left to right) $d=2.99,4.65,7.41\,{\rm kpc}$. 
In each panel, the red strips (that extends from top left to bottom right)
  represent the $50\%$, 
  $90\%$ and $99\%$ confidence intervals
  of the   GW mass function $g(m)$ 
  (similar to the shaded region in the last panel of Fig.~\ref{fig:Acosi-degeneracy}).
The blue strips (that extends from bottom left to top right) 
  represent the   $\pm 5\%,\,10\%$ and $\pm 20\%$ 
  error bars of the binary mass function $f(m)$. 
The black contours represent 2-dimensional confidence intervals
  of $50\%$, $90\%$ and $99\%$. 
The dash-dotted lines represent the true value of
  $f(m)$ and $g(m)$. 
The horizontal and vertical dashed lines represent the true values of 
  $m_{\rm ns}$ and $\cos \iota$-s.
The vertical dotted lines are reference lines of
  $m_{\rm ns}=1.6-2.4 \solarmass$ with $0.1\solarmass$ spacing.}
\label{fig:fmgm}
\end{figure*}

The degeneracy between the GW amplitude $A$ and 
  inclination angle translates into 
  a degeneracy between $m_{\rm NS}$ and $\cos \iota$.
Fig.~\ref{fig:fmgm} illustrates such mass-inclination degeneracy
  from GW and also EM observations for a few 
  NS-WD binaries with SNR$\approx 5$ and $\approx 20$. 
The two almost-diagonal 
  strips represent the mass functions $f(m)$ and 
  $g(m)$, respectively.
The black contours represent the $50\%$, $90\%$, 
  and $99\%$ confidence intervals 
  of 2D PDF of $\cos \iota-m_{\rm ns}$. 
In the following, we will refer to the 
  intersection of $f(m)$ and $g(m)$ strips as $g(m)-f(m)$ constrain
  and the intersection of $f(m)$ strip and 
  2D contour of $\cos \iota-m_{\rm ns}$ from {\tt gbmcmc} simulation
  as ${\rm 2D}-f(m)$ constrain. 


The first column of Fig.~\ref{fig:fmgm} 
  represent the nearly edge-on cases when
  the lower bound of $m_{\rm ns}$ from $f(m)_{\min}$ alone
  out-stands (or is comparable to)
  the joint constraint from the multi-messenger method. 
The upper panel of the first column shows 
  the parameter estimates of  
  a binary system consisting of a $2\solarmass$ NS and 
  a companion star of $m_{\rm c}=0.6\solarmass$ 
  revolving around each other 
  with $P_{\rm orb}=40\,{\rm min}$ at a distance of $5.22\,{\rm kpc}$,
  and viewed at an inclination $\cos \iota=0.1$. 
The system has ${\rm SNR}\approx 5$.   
When $f(m)$ is measured with $\pm 1\%$,
  the $m_{\rm ns} < 1.97\solarmass$ and $m_{\rm ns}>2.17\solarmass$ 
  can be ruled out at 90\% confidence interval
  for ${\rm 2D}-f(m)$ constrain. 
At 50\% confidence interval, the mass is constrained to be
  $1.97-2.08 \solarmass$. 
For a similarly edge-on system with ${\rm SNR}\approx 20$ 
  (lower left panel of Fig.~\ref{fig:fmgm}),
  $\delta_{\rm gm}$ is approximately 
  a quarter of the value in the previous case.
Consequently, the upper bound of $m_{\rm ns}$ from ${\rm 2D}-f(m)$ constrain
  is $2.03 \solarmass$ at $50\%$ confidence interval 
  and $2.04 \solarmass$ at $90\%$ confidence interval 
  which are also about a quarter of the values in the previous case.
Comparing these two cases reveals that the apparent 
  mass-inclination degeneracy from GW observation 
  is evident for edge-on systems at lower SNR, 
  as indicated by the elongated shape of the black contour
  and its alignment with $g(m)$ strips. 
Such degeneracy is absent in the case of edge-on binary at higher SNR,
  in which case the   GW mass function becomes redundant 
  and the NS mass can be determined solely from the GW observations.
Regarding the constraints on $m_{\rm ns}$, 
  the results obtained from $g(m)-f(m)$ constrain
  slightly underestimate the actual constraints 
  derived from the joint GW and EM observations. 

The second column of Fig.~\ref{fig:fmgm} 
  represents cases with intermediate inclination angle $\cos \iota=0.5$.  
The upper middle panel of Fig.~\ref{fig:fmgm} 
  shows a system with ${\rm SNR}\approx 5$. 
If $f(m)$ is measured with an error bar $\pm 1\%$,
  $m_{\rm ns} < 1.6 \solarmass$ and $m_{\rm ns} > 2.5 \solarmass$ can
  be ruled out by both $g(m)-f(m)$ and ${\rm 2D}-f(m)$ 
  at $90\%$ confidence level.
For the lower middle panel of Fig.~\ref{fig:fmgm},
  the SNR is approximately four times the previous case, 
  and the corresponding constraints are
  $1.88-2.10 \solarmass$ for $g(m)-f(m)$ and $1.88-2.12 \solarmass$
  for ${\rm 2D}-f(m)$, respectively,
  when $\delta_{\rm fm}=1\%$. 
The mass-inclination degeneracy from GW observation 
  persists for both cases,
  with $g(m)$ contour aligning nicely with the $m_{\rm NS}-\cos\iota$
  contour for ${\rm SNR}\approx 5$ case,
  and a slight mismatch for ${\rm SNR}\approx 10$ 
  case at small $m_{\rm NS}$,
  which are reminiscent of the first panel of this figure. 
In fact, the mass-inclination degeneracy gradually eases
  when ${\rm SNR} \gtrsim 15$, and further disappear when ${\rm SNR} \gtrsim 40$.

The third column represents nearly face-on 
  NS-WD binaries with $\cos \iota=0.9$. 
For the ${\rm SNR} \sim 5$ case, 
  the 90\% constraints on $m_{\rm ns}$ 
  from ${\rm 2D}-f(m)$ 
  is about $\pm 1\solarmass$.
In contrast, for the case with ${\rm SNR} \sim 20$,
  the constraint narrows to around $\pm 0.24 \solarmass$, 
  and remains insensitive to the value of $\delta_{\rm fm}$.
Both constraints are much worse than 
  those of intermediate inclination angles
  and near $\pi/2$ inclination angles at same SNR. 
For these nearly face-on systems, the mass-inclination degeneracy
  persists even for system with large SNR, which is evident
  from Fig.~\ref{fig:Acosi-degeneracy}.
In fact, within the parameter space considered in this work, 
  this degeneracy consistently persist  for all systems 
  with ${\rm SNR} \le 400$,
  implying its relevance for 
  any realistic GW sources detectable by {LISA}.  
For the selected 6 systems, 
  nearly edge-on systems ($\cos \iota = 0.1$) 
  have tighter constraints on $m_{\rm ns}$ 
  compared to nearly face-on systems 
  ($\cos \iota = 0.9$) when the GW strength is comparable.

\begin{figure*}
\centering
\includegraphics[width=2.15\columnwidth]{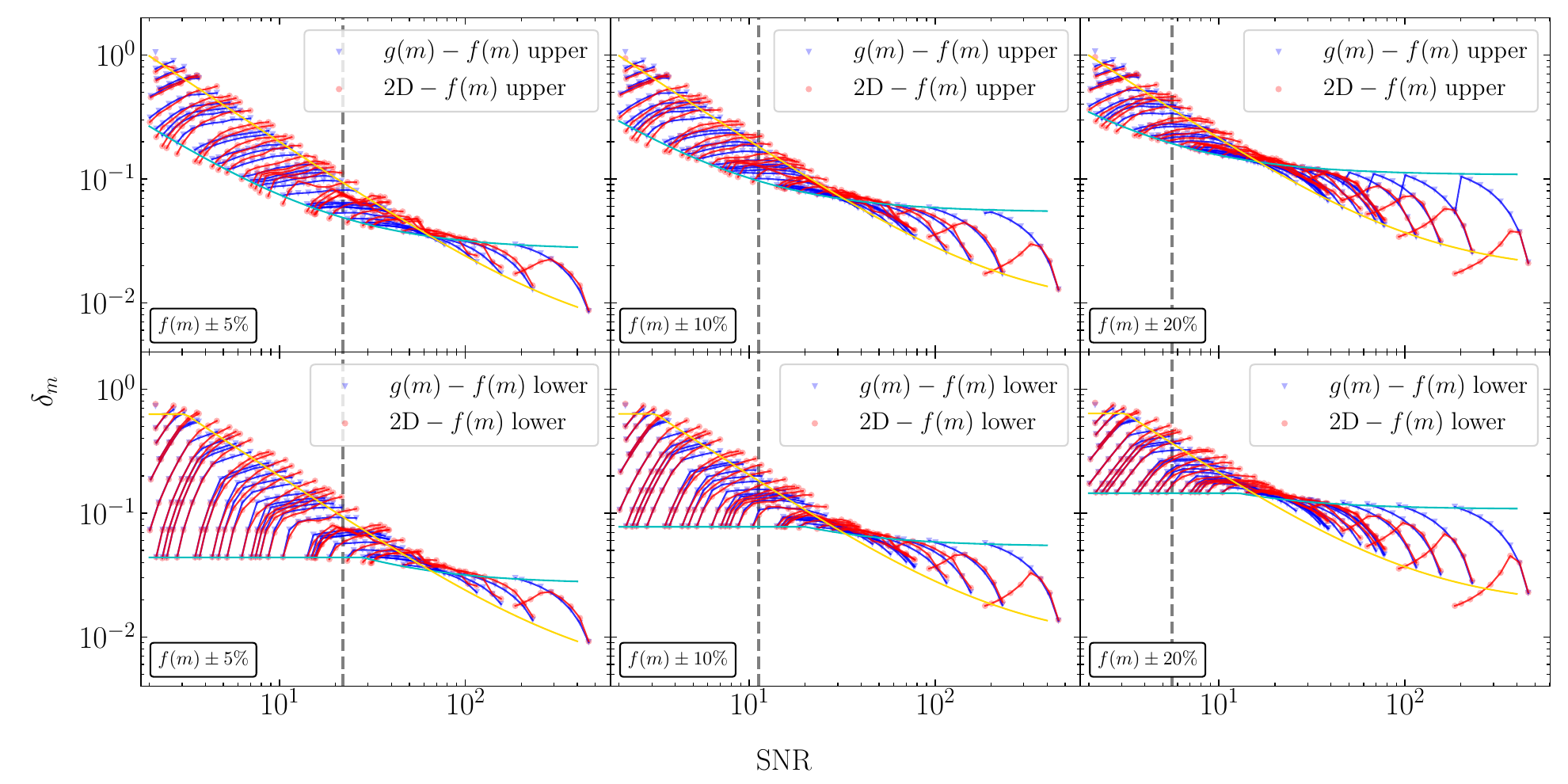}
\caption{
The relative size of the error bar of $m_{\rm ns}$ 
  versus the SNR of a wide range of NS-WD binaries with
  $P_{\rm orb} = 20-60\,{\rm min}$, $\cos \iota=0.1-0.9$ 
  and $d= 1-12\,{\rm kpc}$. 
The upper panels show the 
  upper bound $\delta_{\rm m} \equiv (m_{{\rm ns},\max}-m_{{\rm ns},0})/m_{{\rm ns},0}$, 
  and the lower panels show the lower bound 
  $\delta_{\rm m} \equiv (m_{{\rm ns},0}-m_{{\rm ns},\min})/m_{{\rm ns},0}$.
The $g(m)-f(m)$ and ${\rm 2D}-f(m)$ constraints
  are calculated by finding the intersection of 
  $90\%$ confidence intervals of 1D PDF of $g(m)$ and 
  that of 2D PDF of $\cos\iota-m_{\rm NS}$ with 
  $f(m)\pm \delta_{\rm fm}$,
  assuming   $\delta_{\rm fm}=5\%-20\%$.
The vertical dashed lines represent the reference SNRs 
  when $\delta_{\rm gm} \approx \delta_{\rm fm}$,
  and we have $\delta_{\rm gm} \le \delta_{\rm fm}$ 
  on the left-hand side of the dashed line. 
Systems of the same orbital period 
   and distance, but with different inclination angles 
   are connected with solid lines, with nearly edge-on system 
   (i.e. $\cos \iota=0.1$) on the left end
   and nearly face-on system $\cos \iota=0.9$ on the right end.
The cyan (yellow) line in each panel represents the reference 
  line calculated  
  using Eq.~\ref{eq:mNSError} and Eq.~\ref{eq:gmErrorBar}
  for $\cos \iota = 0.1$ (0.9). 
As depicted in the figure, 
  these reference lines distinctly 
  trace the dependence of $\delta_m$ on SNR
  for systems with the respective inclination angles.
  The turning of the blue curve in the top-right panel 
  for high SNR systems is due to the constraint imposed by $g(m)$ alone (Eq.~\ref{eq:mNSMax_gm}).}
\label{fig:mNSconstrains_Low}
\end{figure*}

Similar parameter estimations were performed for 
  a larger set of binaries with the same $m_{\rm ns}=2\solarmass$, $m_{\rm c}=0.6\solarmass$,
  with different orbital periods $P_{\rm orb}=20-60\,{\rm min}$ located at 
  $d=1-12\,{\rm kpc}$ away, with an inclination angle ranging from $\cos\iota=0.1$ to $\cos\iota=0.9$.
The joint constraints on the NS mass 
  are shown in Fig.~\ref{fig:mNSconstrains_Low}. 
For systems with SNR smaller than the reference SNR, 
  $\delta_{\rm gm} \ge \delta_{\rm fm}$.
In which case, the error bar $\delta_m$ is dominated by
  the uncertainty in GW observation. 
Consequently, as explained in the previous section, 
  error bars for nearly edge-on systems 
  are slightly smaller compared to 
  nearly face-on systems for the same orbital parameters, despite having 
  a slightly larger SNR.
The $g(m)-f(m)$ constraints (blue triangles) align with the 
  joint constraints of ${\rm 2D}-f(m)$ (red circles), 
  particularly for systems 
  with small SNR. 
In this regime, the   GW mass function
  captures the essence of 
  the mass-inclination degeneracy, 
  accurately reproducing and improving upon 
  the joint constraints from multi-messenger observations.

For systems with SNR much larger than the reference SNR, $\delta_{\rm gm} \ll \delta_{\rm fm}$,
  the GW amplitudes of the binary systems 
  are sufficiently large, such that 
  the constraints on NS mass is 
  dominated by the uncertainty of the binary mass function $f(m)$.
In this regime, constraints on $m_{\rm NS}$ for 
  edge-on systems outperform those for face-on systems.
Significant discrepancy   arises
  between $g(m)-f(m)$   constraint
  and ${\rm 2D}-f(m)$   constraint for system with 
  $\cos \iota \le 0.5$ and ${\rm SNR} \ge 100$,
  where the mass-inclination degeneracy has completely eased,
  as shown in the middle and right panels at large SNR in Fig.~\ref{fig:mNSconstrains_Low}.

The lower bounds of $m_{\rm ns}$ for system with $\cos \iota \sim 0$
  are much better than their upper bounds. 
This is due to the stringent constraint from 
  $f(m)_{\min}$ alone. 
When the error bars are much smaller than unity, 
  the lower bounds are approximately 
  $m_{\rm ns} \ge 2\solarmass \times \sin^3 \iota$.

As demonstrated by the systems discussed in this section, 
  the mass constraints derived from the intersection of $g(m)-f(m)$ 
  align closely with those obtained from 
  the intersection of ${\rm 2D}-f(m)$ for NS-WD binaries  
  with ${\rm SNR} \lesssim 100$. 
Moreover, the former is accurately described by
  Eq.~\ref{eq:mNSMax_fmgm} and the empirical relation Eq.~\ref{eq:gmErrorBar}. 
We can therefore conclude that the proposed analytical formula 
  for $\delta_{\rm gm}$, 
  despite being an empirical relation, 
  is well-suited for estimating mass constraints 
  for NS-WD systems across a broad parameter range.

\section{Discussions}
\label{sec:discussion}

\subsection{GW mass-inclination degeneracy}
\label{subsec:apply}



As pointed out in the previous section, 
  the degeneracy between mass and inclination angle
  is prominent in systems with low SNR  
  and it is reduced when SNR increases. 
For nearly edge-on systems with $\cos \iota=0.1$, 
  the degeneracy would become less severe 
  when ${\rm SNR} \gtrsim 20$;  
  for systems with an intermediate inclination  
  of $\cos \iota=0.5$,
  the degeneracy diminishes 
  when ${\rm SNR} \gtrsim 40$. 
However, 
  for nearly face-on systems with $\cos \iota=0.9$,
  the degeneracy persists until ${\rm SNR} \sim 400$.

The origin of this degeneracy lies in the indistinguishability
  of the two polarization modes,
  $h_+$ and $h_{\times}$,
  which depends on both the SNR of the GW signal and 
  the relative strength of the two modes.  
It is reminiscent 
  of the distance-inclination degeneracy
  observed in the analysis of compact object binaries detected by the LIGO-Virgo-KAGRA (LVK) Scientific Collaboration network, 
  which have been studied systematically 
  by \citet{Usman2019}.
These two degeneracies 
  both stem from 
  the degeneracy between GW amplitude 
  and inclination angle (the $A-\iota$ degeneracy), 
  albeit with subtle differences. 

The $A-\iota$ degeneracy can be illustrated  
  with the following example. 
Consider two L-shaped detectors 
  orientated with a $45^{\circ}$ offset,     
  with $\psi=0$ set to 
  correspond to the configuration 
  where the two detectors 
  align with $h_{+}$ and $h_{\times}$ respectively. 
For an exact face-on binary system with 
  an arbitrary polarization angle, 
  the signals recorded 
  in both detectors are identical 
  to those of a closer (or more massive) 
  exact edge-on system with $\psi=\pi/8$ 
  (or $3\pi/8$, etc), 
  with the exception of a phase difference.
An exact edge-on system with 
  polarization angle slightly 
  deviating from $\pi/8$ 
  would therefore resemble 
  a nearly face-on system with an arbitrary polarization angle.
In Bayesian inference, the range of each parameter 
  is determined by
  marginalizing over the remaining parameters. 
Consequently, a nearly edge-on system 
is prone to being
  misconstrued 
  as a nearly face-on system,
  whereas the reverse is not necessarily true. 
The relative strength of two polarization 
    modes $h_{\times}/h_{+} \gtrsim 94\%$ for 
  $\cos\iota \ge 0.7 $, suggesting that
  the GW signal from nearly face-on systems
  become nearly indistinguishable from one another.
This example elucidates  
  the prominent degeneracy observed 
  in Fig.~\ref{fig:Acosi-degeneracy}, 
  and the $\cos \iota=0.5$ and $\cos \iota=0.9$ cases 
  shown in Fig.~\ref{fig:fmgm}.   
Examining the system presented 
  in Fig.~\ref{fig:Acosi-degeneracy},
  where the injection parameter is 
  $\cos \iota = 0.9$, the 
  marginalized posterior of $\cos \iota$ extends towards 
  both smaller and larger values, favoring $\cos \iota=1$,
  as an exact face-on template 
  provides a decent fit to the signal,
  regardless of the polarization angle. 

For a nearly edge-on binary system, 
  the degeneracy is reduced quickly as SNR increases,
  and it disappears 
  when SNR$\gtrsim 20$, 
  a phenomenon also observed in   LVK binaries. 
In   LVK observations, 
  when the the network of GW detectors
  is equally sensitive to both polarization modes, 
  the signals recorded in both detectors 
  are dominated by the $h_{+}$ mode 
  (unless when $\psi$ is very close to $0$ or $\pi/4$). 
The relative strength of the two signals 
  provides additional information about the 
  polarization angle, 
  which will aid in 
  resolving the $A-\iota$ degeneracy. 
  To be more precise, this measurement constrains 
  $\phi - 2\psi$, as demonstrated in 
  Fig.~\ref{fig:Acosi-degeneracy}\footnote{ 
  The case presented here is nearly face-on, 
    but this particular feature also exists for nearly edge-on systems with low SNR. 
  A similar degeneracy was discussed
    by \citet{Usman2019}, although it was reported as $\phi \pm \psi$.
    Our additional factor of $2$ is likely due to a difference in the convention of the polarization angle.}.

In the case of Galactic binaries, 
  the situation is somewhat more intricate. 
Unlike the binaries observed by   LVK network, 
  where the length of the GW signal
  is comparable to the binary
  orbital period 
  (which allows for a precise estimation of $\phi$) 
  the observational time needed for 
  the declaration of a detection   
  is significantly longer than their orbital period 
  for Galactic binaries.
In the latter scenario, 
  information about the initial orbital phase $\phi$ 
  is lost, 
  rendering it as a free orbital parameter.
Due to the strong degeneracy 
  between $\psi$ and $\phi$,
  $\psi$ also   becomes unconstrained. 
The upper panels of Fig.~\ref{fig:fmgm} 
  illustrate 
  such cases. 
Regardless of true inclination angle of the system,
  face-on templates almost always 
  provide a decent fit to the signal,
  suggesting that the parameter estimation results 
  tend to favor, or at least are consistent with 
  larger values of $\cos \iota$. 

For systems with higher SNRs, the situation differs.
As {LISA} rotates in space, 
  the polarization angle varies 
  with respect to the detector's rest frame. 
Consequently, each TDI channel experiences
  at least one episode of high GW signal when 
  its orientation aligns exactly with $h_{+}$
  and one episode of low GW signal when 
  its orientation aligns with $h_{\times}$. 
This distinctive observational feature
  is absent in the low SNR cases 
  due to difficult of 
  signals being entangled with noise. 
Such feature enables us to differentiate
  the amplitude of two polarization modes, 
  thereby constraining the system's inclination angle.
This condition corresponds to the cases 
  in the bottom-left panel of Fig.~\ref{fig:fmgm}.

\subsection{Formation channels of compact NS-WD binaries} 
\label{Existence}

Despite the prevalent belief
\citep{Tauris2018, Korol2024MNRAS, He2024MNRAS} 
  that compact NS-WD systems that are observable with LISA could exist, there is no confirmed observation of such systems. Instead,
  the observed NS-WD binaries are typically found in much wider orbits\footnote{  An up-to-date list of
    observed NS-WD binaries is available in the ATNF catalog: \url{https://www.atnf.csiro.au/research/pulsar/psrcat}}. 
Therefore, it is worth discussing the possible formation channels 
  for the specific binary systems of interest.
NS-WD binaries are evolutionary end points of LMXB/IMXB,  
 and a correlation between the post-LMXB orbital period and 
  the WD mass is expected \citep[see][]{Rappaport1995,Tauris1999,Istrate2014}, 
  where systems with lighter WDs tend to have closer orbits.  
The post-LMXB/post-IMXB evolution 
  is driven 
  mainly by the loss of orbital angular momentum 
  through gravitational radiation 
  \citep[see e.g.][]{Paczynski1971ARA&A}, 
  which constrains that the initial orbital period  
  of the binary has to be $\lesssim 12\;\! {\rm hr}$ 
  in order for it to reach the {LISA} sensitivity band 
  within a Hubble time. 
This narrow range of initial orbital period
  and the correlation of orbital period with WD mass
  at the end of LMXB/IMXB phase suggests that 
  only those with WD mass smaller than $\lesssim 0.2 \solarmass$ 
  can evolve towards this compactness that is relevant for 
  GW observation \citep{Tauris2018}, 
  although the exact WD-mass threshold depends on 
  the mass and metalicity 
  of the progenitor, 
  the efficiency of magnetic braking mechanism
  and the initial orbital period 
  in the pre-LMXB/pre-IMXB phase 
  \citep[see e.g.][]{Tauris1999,Chen2020,Chen2021MNRAS}. 
  
Alternatively, the NS may form  after the WD 
  when the two progenitor stars 
  in the initial binaries are massive enough 
  \citep{Tutukov1993}. 
This formation scenario was proposed 
  to explain the origin
   of PSR B2303+46 and PSR J1141$-$6545 \citep{Tauris2000}, 
  and the resulting binary favors more massive WD,
  i.e. CO or ONeMg WD.


In dense nuclear stellar cluster or globular cluster
  where the stellar population is old, 
  the terminating orbital period of the LMXB phase 
   would be about a factor of two 
   shorter  than those in the field 
   \citep{Rappaport1995}. 
Moreover, the dense stellar environment 
  underscores the significance of dynamical interactions.
Compact NS-WD binaries might be produced  
  through dynamical capture, 
  three-body interactions, or even four-body interactions.
Consider a binary with binding energy 
\begin{align} 
E_{\rm b} & = \frac{G{m_{\rm ns}}^2q}{2a}  
  \nonumber \\ 
  & = \frac{G^{2/3}}{2}\left(\frac{2\pi}{P_{\rm orb}} \right)^{2/3} 
   {m_{\rm ns}}^{5/3}\left[\frac{q}{(1+q)^{1/3}} \right]\ , 
\end{align} 
where $q = m_2/m_{\rm ns}$,  
The hardening rate of it is roughly    
\begin{align} 
\frac{{\rm d}E_{\rm b}}{{\rm d}\;\!t} 
  & \approx   
  \xi  G^2 
  \left(
  \frac{n_* {\bar m}^3}{\langle \sigma \rangle} 
  \right)   
\end{align} 
\citep{Heggie1975MNRAS}, 
 where $\xi \sim 10$, which includes
 contributions to binary hardening 
  via wide/close three-body encounters,
  exchange and resonant encounter\footnote{
  We have dropped the dependency of the results
  on the mass ratio for simplicity, given that all the masses 
  considered here 
  are on the order of unity.}.   
${\bar m}$ represent the averaged mass
 of the encounters,
 $n_*$ represents the stellar number density
 and $\langle \sigma \rangle$ represents the
 velocity dispersion. 
The time scale it takes for 
  a NS-WD binary to harden till the region where
  GW dominates (i.e. when $P_{\rm orb} \sim {\rm days}$) 
  is approximately (assuming $q = 0.4$): 
\begin{align} 
T_{\rm harden} = &  
  \frac{m_{\rm ns}^{5/3}}{2 \xi  G^{4/3}}
  \frac{\left< \sigma \right>}{n_* {\bar m}^3}
  \left(\frac{2\pi}{P_{\rm orb}} \right)^{2/3} 
  \frac{q}{(1+q)^{1/3}} \nonumber \\ 
   \approx & 26  \,{\rm Gyr}\;\!
   \left( \frac{m_{\rm ns}}{2\solarmass} \right)^{{5}/{3}} 
   \left( \frac{10}{\xi} \right)
  \left( \frac{\langle
   \sigma \rangle}
   {200 \;\!{\rm km}~{\rm s}^{-1}}\right)
   \nonumber \\
& \quad\quad 
 \times \left(
   \frac{ 12\;\!{\rm hr}}{P_{\rm orb}}\right)^{{2}/{3}}    
  \left( \frac{10^6 \;\!{\rm pc}^{-3}}{n_*} \right)
  \left( \frac{1 \solarmass}{{\bar m}}\right)^3  \ ,
\end{align}   
which is about double the Hubble time. 
We have adopted a stellar density of 
  $10^6~{\rm M}_\odot\;\!{\rm pc}^{-3}$ 
  at the inner 1~pc \citep{Merritt2010ApJ} of Galactic Centre,
  and a stellar velocity dispersion of $200~{\rm km~s}^{-1}$. 
Similarly, for a typical globular cluster 
   with core stellar density 
  of $\sim 10^4\;\! {\rm M}_\odot\;\!{\rm pc}^{-3}$ \citep{Carballo-Bello2012MNRAS,Pancino2017MNRAS} and velocity
  dispersion of $\sim 5~{\rm km~s}^{-1}$ \citep[see e.g.][]{deBoer2019MNRAS},
  the time scale for hardening is about $5$ times
  the Hubble time\footnote{In the calculations 
    we set ${\bar m}/\solarmass =1$,  
    but stars in the Nuclear Star Cluster 
    of the Galactic Centre 
    \citep[see][]{Chen2023ApJ}
    and in globular clusters 
      are old stars with mass well below 
      $1 \solarmass$. 
Using a more accurate value 
     for ${\bar m}/\solarmass$,
     however, would not alter 
     the conclusion 
     that hardening would take 
     much longer than the Hubble time.}. 
The hardening due to stellar encounters 
  and subsequent GW radiation 
  would therefore 
  be sufficient to produce the target NS-WD binaries
  if the binaries start off sufficiently hard themselves. 
Note that this hardening
  formula has included three-body exchange and resonant encounters,
  both of which tend to eject the least massive star among the triples. 
This process would likely produce 
  NS-WD binary with a heavier 
  WD, the gravitational radiation for which is more efficient.

\section{Conclusions}
\label{sec:conclusion}

Compact NS-WD binaries potentially hold the most massive NSs.
Measuring their masses is 
  essential for our understanding of the densest matter 
  in our Universe. 
In this work, we studied the potential of using GW observation
  to measure the parameters of NS-WD binary and 
  proposed a novel multi-messenger method to constrain NS masses
  in these systems by combining the 
  GW and EM detections. 
For a series of hypothetical NS-WD binaries 
  with sub-hourly orbital period inside our Galaxy, 
  we generated $4$ years of mock GW data for each system and performed
  parameter estimation using the {\tt gbmcmc} package.  
By combining the GW information with the binary mass function from EM observation,
  the degeneracy between mass and inclination angle can be resolved,
  allowing for much tighter constraints on the NS mass. 
We showed that for our multi-messenger method, 
  the major factors that determine 
  the constraint of the NS mass are
  the GW SNR of the NS-WD binary, the accuracy of the binary mass function, 
  and the inclination angle. 
In general, for a realistic range of SNR (i.e. $\le 100$), 
  edge-on systems are favored over face-on systems in terms of
  parameter estimation.

Regardless of the inclination of the systems, 
  our method suggests that the $m_{\rm ns}$ of a NS-WD binary
  can be constrained 
  to within $\pm 0.2\solarmass$ as long as their   ${\rm SNR} \ge 25$,
  and the binary mass function is measured within $\pm 10\%$ accuracy. 
For a NS-WD binary with $P_{\rm orb} \sim 60\,{\rm min}$ and 
  $m_{\rm c} = 0.6\solarmass$, the system 
  needs to be within $\sim 1\,{\rm kpc}$.
However, for a system with $P_{\rm orb} \sim 20\,{\rm min}$,
  the distance threshold for the acceptable mass determination is about 
  $\sim 20 \,{\rm kpc}$.  
Our study suggests that those NS-WD binaries in globular clusters and in
  our Galaxy can be used to constrain the mass of massive NS.


Last but not least, 
  we defined a new concept:   GW mass function $g(m)$, which is derived
  from GW observations, in contrast to the binary mass function
  derived from optical photometric or spectroscopy observations. 
We showed that the error bar of the   GW mass function
  follows a simple power law relation 
  with the GW SNR of NS-WD binaries,
  regardless of the companion's mass,
  the orbital period,
  or the viewing inclination, 
  validating the universality of this definition.
Moreover, the   GW mass function captures the essence of the 
  mass-inclination angle degeneracy in GW observations,
  and therefore, serves as a handy tool for constraining $m_{\rm ns}$ with multi-messenger observations.


\section*{Acknowledgements}

We thank the anonymous referee  
  of a previous paper of ours 
  for encouraging us 
  to fully develop the conceptual framework  
  of   the GW mass function. 
We thank Silvia Zane for the discussions on 
      mass-radius relations of NSs 
      and various tests in future X-ray studies,  
  Daisuke Kawata 
     on stellar populations and density 
     at the Galactic Centre,  
  and Qin Han on the evolution of compact binaries.  
We also thank Jane Yap, Joana Teixeira and Jun Lau
  for general discussions and comments on this work. 
KW thanks the hospitality of NTHU IoA, 
  where a substantial part of this work was conducted,  
  during his visits. 
KJL is supported by a PhD Scholarship from the 
  Vinson and Cissy Chu Foundation 
  and by a UCL MAPS Dean's Prize. 
KW and JSL acknowledge the support 
  of the UCL Cosmoparticle Initiative. 
This work is supported 
   by the National Science and Technology Council of Taiwan (ROC) 
   under the grants 110-2628-M-007-005 and 112-2112-M-007-042 (PI: A.~Kong). 
This work has made use of the NASA Astrophysics Data System. 



\software{GetDist \citep{Lewis2019arXiv}, Astropy \citep{astropy:2022}, gbmcmc \citep{ldasoft2020}, 
  Mathematica \citep{Mathematica12}}


\bibliography{GW}{}
\bibliographystyle{aasjournal}




\appendix

\section{Remarks on Assumptions and Error propagation}
\label{app:A}

The purpose of this study 
 is to demonstrate the 
 the potential of utilizing GW observations 
  to constrain NS mass in conjunction with EM observations.  
Therefore, we have not incorporated uncertainties 
  in the distance to the system and in the mass of the WD.
For a more rigorous analysis, these uncertainties 
  must be explicitly included in the error propagation.

For WDs are sufficient bright in the optical wavebands,  
  {Gaia} can provide accurate distance measurements. 
To determine the mass of a WD, 
  as an isolated source in the field 
  or as a component star in a binary,  
  is less straightforward. 
The most widely adopted method for WD mass determination 
  relies on the measurement of surface gravity 
  and effective temperature 
  by fitting spectroscopic data 
  with atmospheric models \citep[see e.g.][]{Holberg1985,Holberg1986,Bergeron1992,Bragaglia1995,Schmidt1996,Finley1997,Barstow2001,Kepler2017}. 
The results, combined with a theoretical 
  mass-radius relation for finite-temperature WD structure \citep[see e.g.][]{Althaus2005,Renedo2010,Romero2015}, enable reasonably accurate mass inferences. 
A complication is that 
  whether the WD in NS-WD binaries 
  would adhere to this evolutionary mass-radius relation,
    particularly in the presence 
  of tidal effects and potential pulsar irradiation.
However, 
  when an accurate distance measurement is available, 
  combining WD luminosity 
  with effective temperature information 
  allows deducing radius and, 
  consequently, mass without 
  detailed modelling of atmospheric physics \citep{Koester1979,Bedard2017,Genest-Beaulieu2019}. 
Conversely, this method can also be used 
  to derive the distance to the binary 
  when combined with surface gravity measurement 
  if an accurate parallax measurement is unavailable.
If the WD undergoes a certain evolutionary phase, 
  pulsation mode analysis provides accurate mass measurements
  \citep[see e.g.][]{Kawaler1990,Bradley1994,Corsico2019}. 

While a detailed analysis of the accuracy and systematics
  in mass and distance determination 
  is beyond the scope of this work,
  we still include here 
  a recipe for error propagation  
  appropriate for applications in real observations.  
When the error bars are much smaller than unity,
  we can write $d = d_0 (1 \pm \delta_{\rm d})$,
  where $d_0$ represents the true value of distance
  and $\delta_{\rm d}$ represents the relative size of its error bar. 
The error of distance 
  enters the   GW mass function via  
\begin{align}
\delta_{\rm gm, re} & = 
\delta_{\rm gm} + \frac{3}{5}\;\! \delta_{\rm d} \ ,
\end{align}     
where $\delta_{\rm gm, re}$ represents the realistic 
  relative error bar of $g(m)$,
  and $\delta_{\rm gm}$ represents the case with known distance 
  (as used in the main text of this paper).
Therefore, a $10\%$ uncertainty of the distance
  translates into a $10\%$ uncertainty of $m_{\rm ns}$, approximately.

The companion mass does not enter the
  GW parameter estimation process, 
  and therefore it does not affect the derived uncertainty of   the GW mass function even if it is unknown. 
However, its uncertainty 
  will affect the  
  constraint 
  on the NS's mass, 
  as its value is used  to derive
  $m_{\rm ns}$ from the intersection of two mass functions.
We suppose 
  that the companion's mass 
  $m_{\rm c} = m_{{\rm c},0} (1 \pm \delta_{\rm c})$.
  
When all system parameters are measured with errors 
  much smaller than unity, 
  Eq.~\ref{eq:mNSMax_fmgm} can be approximately solved 
  using series expansion. 
Keeping all errors to the leading order, the relative error 
  on the NS mass can be approximated as:
\begin{align} 
\label{eq:mNSError}
\delta_{\rm m} & =  C_{\rm gm} \delta_{\rm gm,re} + C_{\rm fm} \delta_{\rm fm} + C_{\rm c} \delta_{\rm c} \ , \nonumber \\ 
& = 
\frac{1}{32 (\cos 2 \iota_0 +3) m_{{\rm ns},0}-3 m_{{\rm c},0} 
(-4 \cos 2 \iota_0 +\cos 4 \iota_0 -61)} \nonumber \\ 
& \hspace*{0.5cm} \times  
 \bigg\{
 8  \sin^2 \iota_0  (\cos 2 \iota_0 +7) (m_{{\rm c},0}+m_{{\rm ns},0} 
  )\delta_{\rm fm}  \nonumber \\
& \hspace*{0.8cm} +5 (28 \cos 2 \iota_0 +\cos 4 \iota_0 +35) (m_{{\rm c},0}+m_{{\rm ns},0} )   \left(\delta_{\rm gm} + \frac{3}{5}\;\! \delta_{\rm d} \right) 
  \nonumber \\
& \hspace*{0.8cm} - \big[
2 m_{{\rm c},0} \left(52 \cos 2 \iota_0+3 \cos 4 \iota_0+9\right) + 3 (28 \cos 2 \iota_0 
\nonumber \\
& \hspace*{1.5cm} +\cos 4 \iota_0  +35) m_{{\rm ns},0}
\big]\delta_{\rm c}
\bigg\} \  , 
\end{align}     
where $ C_{\rm gm} \delta_{\rm gm,re}$ 
  and $C_{\rm fm} \delta_{\rm fm}$ 
  are the contribution from $f(m)$ and $g(m)$,
  and $C_{\rm c} \delta_{\rm c}$ is the contribution from 
  $m_{\rm c}$.
The coefficient $C_{\rm c} \in [-0.28,0]$ 
  for edge-on system and 
 $\sim -1.3$ for face-on system when 
  $m_{{\rm c},0}=[0.1,0.6]\solarmass$.
Note that the orders of magnitude of 
  all coefficients ($C$) are unity, 
  indicating that the uncertainty in 
  $m_{\rm ns}$ will be dominated by the measurement 
  with the greatest uncertainty $\delta$. 
To utilize this method 
  for the precise determination of NS mass, 
  accurate measurements of distance and the WD mass 
  are necessary.


\label{lastpage}
\end{document}